\documentclass[12pt,A4paper]{article}
\usepackage{graphicx}
\usepackage{amsmath}
\usepackage{amssymb}
\usepackage{cite}
   
\newcommand\Ref[1] {Ref.\,\cite{#1}}

\newcommand\eqn[1] {Eq.\,(\ref{#1})}

\newcommand\fig[1] {Fig.\,{\ref{#1}}}

\def\beq{\begin{equation}}
\def\eeq{\end{equation}}
\def\beqq{\begin{equation*}}
\def\eeqq{\end{equation*}}
\def\bsp#1\esp{\begin{split}#1\end{split}}
\def\bal#1\eal{\begin{align}#1\end{align}}

\newcommand\ud   {\mathrm{d}}

\newcommand{\baikovmeasure}[1]{\ensuremath{\frac{\prod \ud x_i}{\prod_{{#1}}x_i^{\alpha_{i}}}}}

\begin{document}

%%%%%%%%%%%%%%%%%%%%%%%%%%%%%%%%%%%%%%%%%%%%%%%%%%%%%%%%%%%%%%%%%%%%%%
%%%%%%%%%%%%%%%%%%%%%%%%%%%%%%%%%%%%%%%%%%%%%%%%%%%%%%%%%%%%%%%%%%%%%%
%%% 
%%% The Title Page
%%%
%%%%%%%%%%%%%%%%%%%%%%%%%%%%%%%%%%%%%%%%%%%%%%%%%%%%%%%%%%%%%%%%%%%%%%
%%%%%%%%%%%%%%%%%%%%%%%%%%%%%%%%%%%%%%%%%%%%%%%%%%%%%%%%%%%%%%%%%%%%%%

\begin{titlepage}
\renewcommand{\thefootnote}{\arabic{footnote}}
\par \vspace{2cm}

\begin{center}
{\Large \bf  A new reduction strategy for special negative
sectors of planar two-loop integrals without Laporta algorithm}
\end{center}
\par \vspace{10mm}
\begin{center}
{\bf Adam Kardos}

\vspace{0mm}

University of Debrecen, Faculty of Science and Technology, Institute of Physics\\
H-4010 Debrecen, PO Box 105, Hungary
\end{center}

\par \vspace{2cm}
\begin{center} {\large \bf Abstract} \end{center}
\begin{quote}
\pretolerance 10000
In planar two-loop integrals there is a dedicated sector such that when its index is zero,
the two-loop integral decomposes into the product of two one-loop integrals. We show an 
alternative reduction strategy for these sectors when their index is negative
using the Baikov representation. This reduction strategy is free from the Laporta algorithm. 
It follows a top-down approach and is much faster than approaches based on the brute-force, 
conventional integration by parts identities.
\end{quote}

\vspace*{\fill}
\begin{flushleft}
December 2018
\end{flushleft}
\end{titlepage}
\clearpage

\renewcommand{\thefootnote}{\fnsymbol{footnote}}

%%%%%%%%%%%%%%%%%%%%%%%%%%%%%%%%%%%%%%%%%%%%%%%%%%%%%%%%%%%%%%%%%%%%%%
%%%%%%%%%%%%%%%%%%%%%%%%%%%%%%%%%%%%%%%%%%%%%%%%%%%%%%%%%%%%%%%%%%%%%%
%%%
%%% Body of Text
%%%
%%%%%%%%%%%%%%%%%%%%%%%%%%%%%%%%%%%%%%%%%%%%%%%%%%%%%%%%%%%%%%%%%%%%%%
%%%%%%%%%%%%%%%%%%%%%%%%%%%%%%%%%%%%%%%%%%%%%%%%%%%%%%%%%%%%%%%%%%%%%%

The extremely successful operation of LHC resulted in high-quality data taken by all its 
experiments. This data is being used to produce high-accuracy measurements for several
processes to stress-test the standard model. From the theory side equally high-precision 
predictions are required to be able to draw definite conclusions from these comparisons.
After the full automation of NLO calculations in QCD \cite{Gleisberg:2008ta,Bevilacqua:2011xh,Alwall:2014hca}. Tools are becoming available to perform NNLO calculations 
\cite{Catani:2008me,Catani:2009sm,Gavin:2012sy,Ridder:2014wza,Boughezal:2016wmq,	Grazzini:2017mhc,Currie:2018oxh} and automation at NNLO is put on the horizon. One crucial 
ingredient of these calculations is the two-loop amplitude.

In the two-loop amplitude the occurring tensor integrals have to be expressed on an integral
basis. One way to do so is by means of integration by part (IBP) identities \cite{Tkachov:1981wb,Chetyrkin:1981qh} by which these tensor integrals are written as a linear combination of master
integrals times coefficients depending on kinematic invariants and space-time dimension.
In practice Laporta's algorithm \cite{Laporta:2001dd} is used to derive IBP identities for 
reduction. This algorithm is implemented in several computer programs to tackle the problem of
reduction \cite{Anastasiou:2004vj,vonManteuffel:2012np,Lee:2012cn,Smirnov:2014hma,Maierhoefer:2017hyi}.
As for the computation of remaining master integrals the last decade witnessed an unprecedented 
advancement.
This spans from how to write these master integrals as differential equations \cite{Henn:2013pwa,Papadopoulos:2014lla,Bosma:2017hrk} through the definition of new functions
\cite{Goncharov:1998kja,Goncharov:2001iea,Goncharov:2010jf,Ablinger:2013cf,Broedel:2018iwv}
to new techniques and tools to attack the resulting equations 
\cite{Lee:2014ioa,Gituliar:2017vzm}.

These techniques are put to the ultimate test when two-loop five-parton amplitudes were
calculated \cite{Badger:2015lda,Gehrmann:2015bfy,Chawdhry:2018awn,Badger:2018enw,Abreu:2018zmy}. In a realistic calculation due to the numerator
 structure 
propagators not only appear with positive but also with negative and large exponents. As it was found
in Ref. \cite{Chawdhry:2018awn} the real bottle-neck for reduction came from those integrals 
where propagators had a large negative exponent. 

At the two-loop level planar integrals are special
in the sense that they only contain one propagator depending on both loop momenta and in the absence
of this propagator the two-loop integral can be written as the product
of two one-loop tensor integrals. Because of this the sector associated to this propagator is treated
as a special one. If this nature of the special sector is not recognized and the reduction is
carried out as an ordinary one the reduction time can be unnecessarily large.

In this work we present an alternative reduction strategy for this special sector of two-loop
planar integrals. In the heart of this reduction strategy stands
the Baikov representation \cite{Baikov:1996iu,Baikov:1996rk,Smirnov:2003kc,Lee:2010wea,Grozin:2011mt,Lee:2013hzt,Harley:2017qut}. In general it was found that a reduction strategy using the Baikov 
representation results in very cumbersome systems of equations \cite{Bosma:2017hrk}. However,  for this special 
sector the Baikov representation offers a nice way to perform the reduction by disentangling the 
two-loop integral into
two one-loop integrals. The resulting one-loop tensor integrals are much easier to run an IBP-based or
Passarino-Veltman reduction on them. The special sector
having a negative index the original two-loop integral can also be considered as the product of two
one-loop tensor integrals on which a Passarino-Veltman reduction can in principle be done. Nonetheless
with our reduction strategy we can offer one more option laying between the blind IBP and traditional
Passarino-Veltman reduction.

In order to derive our reduction strategy for the special sector we first introduce the Baikov
representation for a general $L$-loop Feynman integral. In this effort we closely follow the 
notation of \Ref{Frellesvig:2017aai}.
An $L$-loop Feynman-integral with $E$ independent external legs the integral can be
written as
\begin{align}
I^{(L)}_{\alpha_1\dots\alpha_N} &= \int\left(\prod_{i=1}^{L}\frac{\ud^d \ell_i}{i\pi^{d/2}}\right)
\frac{1}{D_1^{\alpha_1}\cdots D_N^{\alpha_N}}
\,,
\end{align}
where $N = \frac{L(L+1)}{2} + L E$, $\alpha_i$ are the positive or negative integer
indices and $D_i$ are the propagator factors. Every propagator factor can be written in the form 
of
\begin{align}
D_a &= \sum_{i,j=1}^{L}A_a^{ij}(\ell_{i}\cdot \ell_{j}) +
\sum_{i=1}^{L}\sum_{j=1}^{E}A_{a}^{i(j+L)}(\ell_{i}\cdot p_{j}) + f_{a}
\,,\quad a\in{1,\dots,N}
\end{align}
where $A_{a}^{ij}$ are the coefficient matrices and $f_a$ comprise the dot products involving  
only external momenta and internal masses. The definition of coefficient matrices and the $f_a$ 
functions allow us to write the multiloop Feynman integral in the form of Baikov:
\begin{align}
I^{(L)}_{\alpha_1\dots\alpha_N} &= \mathcal{N}\int\frac{\ud x_1\cdots \ud x_{N}}
{x_{1}^{\alpha_1}\cdots x_{N}^{\alpha_{N}}}
\left(\mathcal{P}_{N}^{L}(x_1 - f_1,\dots,x_N - f_N)\right)^{\frac{d-L-E-1}{2}}
\label{eqn:BaikovDef}
\end{align}
with
\begin{align}
\mathcal{P}_{N}^{L}(x_1,\dots,x_{N}) &= 
\left.G(\ell_1,\dots,\ell_L,p_1,\dots,p_{E})\right|_{s_{ij} = \sum_{a=1}^{N}A_{a}^{ij}x_{a}}
\,,
\end{align}
where $G$ is the Gram determinant composed of momenta appearing in its argument, the $x$'s
are called the Baikov variables and $\mathcal{N}$
is a factor depending on the topology, dimension and external momenta -- not relevant to the
forthcoming discussion -- the exact definition can be found, e.g., in Ref. \cite{Frellesvig:2017aai}.

\begin{figure}[!t]
\centering
\includegraphics[width=6cm]{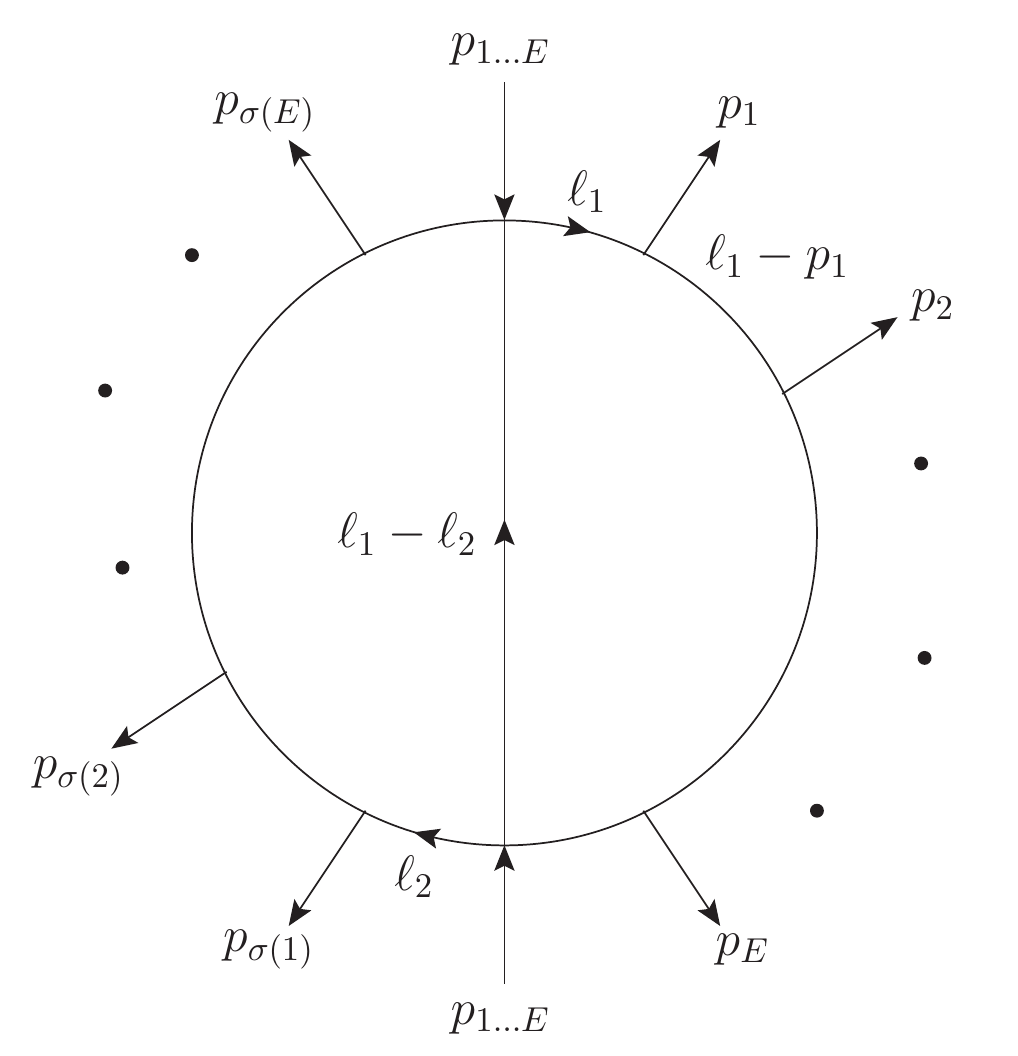}
\caption{{\label{fig:general_planar}} A general two-loop planar topology with $E+1$ external legs and
$\sigma\in S_{E}$.}
\end{figure}

Due to construction the Baikov polynomial is at most quadratic in each Baikov variable. In a 
generic
planar two-loop Feynman integral it is always possible to have only one propagator containing both 
loop momenta. Such a Feynman integral is depicted on \fig{fig:general_planar}. The Baikov polynomial
is the determinant of the Gram matrix where we substitute linear combinations of $x$'s for the 
dot products. If using $q_i\in\{\ell_1,\dots,\ell_L,p_1,\dots,p_E\},\,i\in\{1,\dots,L+E\}$ the
Gram determinant can be written as:
\begin{align}
G(\ell_1,\dots,\ell_{L},p_1,\dots,p_E) &=
\sum_{\sigma\in S_{L+E}}\left(\mathrm{sgn}(\sigma)\prod_{i=1}^{L+E}q_{i}\cdot q_{\sigma_i}\right)
\,,
\label{eqn:gramdetdef}
\end{align}
which is just the definition of the determinant for a $(L+E)\times (L+E)$ matrix where each matrix
element is a genuine dot product and $S_{L+E}$ is the set containing all permutations of the
$L+E$ elements. It is apparent from \eqn{eqn:gramdetdef} that when $L=2$ the
terms proportional to $(\ell_1\cdot \ell_2)^2$ beside of this factor can only contain dot products
of external momenta. Thus for a two-loop planar topology if $x_a$ assigned to the propagator 
containing both loop momenta in the  corresponding Baikov polynomial the terms having $x_a^2$ 
dependence beside this factor can only contain external invariants or internal masses. Because of the
quadratic dependence on Baikov variables, the Baikov polynomial can be written as
\begin{align}
\mathcal{P}^{L}_{N}(x_1,\dots,x_{N}) = f(x_1,\dots,x_{a-1},x_{a+1},\dots,x_N)
(x_a^{+} - x_a)(x_a - x_a^{-})
\,,
\end{align}
the integration region for $x_a$ extends from $x_a^{-}$ to $x_a^{+}$ and with the obvious, gracious
property of 
$\left.\mathcal{P}_{N}^{L}(x_1,\dots,x_{N})\right|_{x_a = x_a^{-}} = 
\left.\mathcal{P}_{N}^{L}(x_1,\dots,x_{N})\right|_{x_a = x_a^{+}} = 0$.
To simplify our notation we define our multiloop Feynman integral in terms of the Baikov variables
such that
\begin{align}
I^{(L)}_{\alpha_1\dots\alpha_N} &= \mathcal{N}
\int\baikovmeasure{}\mathcal{P}^n
\,,
\end{align}
where $\mathcal{N}$ is the factor already used in \eqn{eqn:BaikovDef},  $n = \frac{d - L - E - 1}{2}$ 
and for brevity we dropped both sub- and superscript and argument of the Baikov polynomial, 
$\mathcal{P}$. We are interested in the two-loop planar case where the special sector, the
one having both loop momenta, has a negative exponent, can be written as:
\begin{align}
I^{(2)}_{\alpha_1\dots\alpha_N} &=
\mathcal{N}\int\baikovmeasure{i\ne a} x_a^{-\alpha_{a}}\mathcal{P}^n
\,, \quad \alpha_a < 0\,.
\end{align}
First, let us examine the case where $\alpha_a = -1$. We know that the coefficient of the term
proportional to $x_a^2$ in the Baikov polynomial does not depend on other $x$'s thus the 
coefficient of the term proportional to $x_a$ in $\partial_{x_a}\mathcal{P}$ cannot have any 
dependence on $x$'s either. Hence by substituting $\partial_{x_a}\mathcal{P}$ for $x_a$:
\begin{align}
\mathcal{N}\int\baikovmeasure{i\ne a} (\partial_a\mathcal{P})\mathcal{P}^n
&=
\mathcal{C}_{\alpha_1\dots\alpha_N}I^{(2)}_{\alpha_1\dots\alpha_N} + 
\sum_{\substack{\{\beta\}\\ \beta_a = 0}}
\mathcal{C}_{\beta_1\dots\beta_N}I^{(2)}_{\beta_1\dots\beta_N}
=
\nonumber\\
&=
\mathcal{C}_{\alpha_1\dots\alpha_N}I^{(2)}_{\alpha_1\dots\alpha_N} + 
\sum_{\substack{\{\beta\}\\ \beta_a = 0}}\mathcal{C}_{\beta_1\dots\beta_N}
\left(I^{(1)}\otimes I^{(1)}\right)_{\beta_1\dots\beta_N}
\,,
\end{align}
where as a short-hand $\partial_a$ was used for $\partial_{x_a}$, $\{\beta\}$ stands for a
set of indices. The source of these extra index configurations is the polynomial nature of
$\partial_a\mathcal{P}$ which does not only contain $x_a$ with some prefactors but
additional terms too. The additional terms can depend on several Baikov variables and hence
can alter the power these variables appear on.
The $\mathcal{C}$ coefficients only depend on external kinematics and masses. 
As the special sector
appears on the zeroth power in all the terms in the right hand side but the first one and
this is the only sector containing both loop momenta these integrals are products of two one-loop
Feynman integrals, hence the notation $I^{(1)}\otimes I^{(1)}$. In order to use this observation in
the reduction of two-loop planar integrals we have to turn this into an IBP relation. To this end note
that the integration limits for all the Baikov variables are determined by the condition 
$\mathcal{P} = 0$. Keeping this in mind we can come up with the following relation:
\begin{align}
\int\baikovmeasure{i\ne a} (\partial_a\mathcal{P})\mathcal{P}^n &=
\frac{1}{n+1}\int\baikovmeasure{i\ne a}\partial_{a}\left(\mathcal{P}^{n+1}\right)
=
\nonumber\\
&=\widetilde{\mathcal{C}}_{\alpha_1\dots\alpha_N}I^{(2)}_{\alpha_1\dots\alpha_N} + 
\sum_{\substack{\{\beta\}\\ \beta_a = 0}}\widetilde{\mathcal{C}}_{\beta_1\dots\beta_N}
\left(I^{(1)}\otimes I^{(1)}\right)_{\beta_1\dots\beta_N} = 0
\,,
\label{eqn:SpecialMaster}
\end{align}
where the reader should notice that the integration over $x_a$ only results in vanishing surface 
terms since the only dependence on $x_a$ in the integrand is in the Baikov polynomial in the second
step. The tilde over the coefficients is introduced because for simplicity we divided
\eqn{eqn:SpecialMaster} by $\mathcal{N}$, $\widetilde{\mathcal{C}} = \mathcal{C} / \mathcal{N}$.

This way we can always write a two-loop planar Feynman integral having the special inverse
propagator on the first power as a sum of products of one-loop integrals. To do so we do not 
have to perform the full reduction, instead, we can start with the special sector. Upon reduction
the original integral becomes a product of two one-loop tensor integrals for which the reductions 
are much simple.

So far we only addressed the case when the exponent of the special sector is $-1$. To have a
useful alternate reduction strategy for this kind of integrals we have to devise a strategy even for
the case when $\alpha_a < -1$ as well. If the propagator corresponding to the special sector is 
raised to the power $\alpha_a$, due to the properties outlined previously of the Baikov 
polynomial, the following relation holds:
\begin{align}
\int\baikovmeasure{i\ne a} 
x_{a}^{-\alpha_{a}-1}(\partial_a\mathcal{P})\mathcal{P}^n &=
\widetilde{\mathcal{C}}_{\alpha_1\dots\alpha_N}I^{(2)}_{\alpha_1\dots\alpha_N} +
\sum_{\substack{\{\beta\}\\ \alpha_a < \beta_a}}
\widetilde{\mathcal{C}}_{\beta_1\dots\beta_N}
I^{(2)}_{\beta_1\dots\beta_N}
\,,
\end{align}
where the first term on the right-hand side is the integral we want to reduce, with some 
prefactors, and the remaining terms are further two-loop integrals where the special sector
appears on a lower power.

This relation can be turned into an IBP identity by recasting it into:
\begin{align}
\int\baikovmeasure{i\ne a} 
x_{a}^{-\alpha_{a}-1}(\partial_a\mathcal{P})\mathcal{P}^n &=
\frac{1}{n+1}\int\baikovmeasure{i\ne a}
\partial_a\left(x_{a}^{-\alpha_{a}-1}\mathcal{P}^{n+1}\right)
+\nonumber\\
&+ \frac{1+\alpha_{a}}{n+1}\int\baikovmeasure{i\ne a}x_{a}^{-\alpha_{a}-2}\mathcal{P}^{n+1}
\,,
\end{align}
where the first term on the right-hand side is zero because it is a total derivative in $x_{a}$.
Thus we find that:
\begin{align}
0 &= 
\int\baikovmeasure{i\ne a} 
x_{a}^{-\alpha_{a}-1}(\partial_a\mathcal{P})\mathcal{P}^n 
- \frac{1+\alpha_{a}}{n+1}\int\baikovmeasure{i\ne a}x_{a}^{-\alpha_{a}-2}\mathcal{P}\,\mathcal{P}^{n}
\,,
\end{align}
where both $x_{a}^{-\alpha_{a}-1}\partial_a\mathcal{P}$ and $x_{a}^{-\alpha_a-2}\mathcal{P}$ have
terms proportional to $x_{a}^{-\alpha_{a}}$. Collecting these terms we notice that the rest have 
the Baikov variable $x_a$ on a lower power thus our original integral can be expressed through
ones having lower rank in $x_a$. Notice also that the special case immediately follows from setting 
$\alpha_{a} = -1$.

Reductions can be further simplified when the Baikov polynomial is written\footnote{In a practical
implementation of the algorithm in \texttt{Mathematica} we used the built-in function called
\texttt{PolynomialReduce} for this purpose.} in the form of:
\begin{align}
\mathcal{P} &= \sum_{j=1}^{N}g_j\frac{\partial\mathcal{P}}{\partial x_{j}} + b
\label{eqn:BaikovDecomp}
\,,
\end{align}
where both $g_j$ and $b$ are polynomials in the Baikov variables. Thus the IBP relation for the
general case can be cast into the form of:
\begin{align}
0 &= 
\int\baikovmeasure{i\ne a} 
x_{a}^{-\alpha_{a}-1}(\partial_a\mathcal{P})\mathcal{P}^n 
- \frac{1+\alpha_{a}}{n+1}
\left\{
\sum_{j=1}^{N}
\int\baikovmeasure{i\ne a}x_{a}^{-\alpha_{a}-2} g_{j}(\partial_{j}\mathcal{P})\mathcal{P}^{n}
+
\right.
\nonumber\\
&+
\left.
\int\baikovmeasure{i\ne a}x_{a}^{-\alpha_{a}-2} b\,\mathcal{P}^{n}
\right\}\,.
\label{eqn:GeneralIBP}
\end{align}
This form turns out to be very useful
since it can be used to further simplify the reduction of a given integral. For some $j$ the
monomial coefficient of $(\partial_j \mathcal{P})\mathcal{P}^n$ in the integrand can become
independent of $x_j$ thus the term can be dropped being a total derivative in $x_j$.

In practice when given a planar two-loop integral with the mixed propagator on the $\alpha_a$ 
power we calculate the Baikov polynomial for the integral family and the decomposition of 
\eqn{eqn:BaikovDecomp}. Plugging these into \eqn{eqn:GeneralIBP} with $\alpha_{a}$ and 
$n = \frac{d-L-E-1}{2}$ we can identify the integral we started with and it can be expressed with
integrals having the special sector on a lower power. This procedure can be continued until we
have only one integral having the special sector on the minus first power and all the other
integrals are just products of one-loop tensor integrals. Solving for this integral and 
substituting back we performed the reduction of our original two-loop tensor integral
in terms of products of one-loop tensor integrals. So the reduction of the integral follows
a simple top-down strategy, free from the Laporta algorithm.

In this letter we showed an alternate reduction strategy for a sector of planar two-loop 
tensor integrals which contains both loop momenta and the corresponding inverse propagator 
appears in the numerator with some positive power. The basis of the reduction is the Baikov
representation of the integral and the exploitation of basic properties of this representation.
As the reduction is carried out, the original two-loop tensor integral becomes a sum of 
products of two one-loop tensor integrals. We tested our reduction strategy with
Laporta-based reduction programs available in the literature and we found agreement for all, including  
the most complicated topologies, like the pentabox and massive doublebox. 
We found that with a non-optimized \texttt{Mathematica}
implementation of our strategy it was possible to carry out the 
reduction even for tensor integrals with high rank in the mixed sector on a single laptop. On the
other hand using commercially available software we had to use a rack-mounted computer with 48
cores equipped with significant amount of memory to perform the same operation. In an NNLO
calculation we encounter both two-loop amplitudes and interference terms of two
one-loop amplitudes. The reduction of the former to master integrals is a tedious
procedure but with our method applied to the special sector of the planar part a significant 
amount of time can be saved. In case of the latter the appearance of the special sector is
natural. The presence of these contributions makes it cumbersome to use numerical routines to
evaluate the tensor integrals. By applying our prescription to the problem the mixed 
inverse propagator disappears and the contribution becomes the product of two genuine one-loop 
tensor integrals attackable even with numerical programs dedicated to one-loop amplitude
reductions.

The author is grateful to Costas Papadopoulos for so many fruitful discussions on the topic and
for his comments on the manuscript, Zolt\'an Tr\'ocs\'anyi for carefully reading the manuscript 
and Manfred Kraus for providing far-from trivial two-loop planar
tensor integrals to test the scheme. We acknowledge financial support from the Premium 
Postdoctoral Fellowship program of the Hungarian  Academy  of  Sciences.   This  work  was 
supported  by  grant  K  125105  of  the National Research, Development and Innovation Fund in 
Hungary.

%%%%%%%%%%%%%%%%%%%%%%%%%%%%%%%%%%%%%%%%%%%%%%%%%%%%%%%%%%%%%%%%%%%%%%
%%%%%%%%%%%%%%%%%%%%%%%%%%%%%%%%%%%%%%%%%%%%%%%%%%%%%%%%%%%%%%%%%%%%%%
%%%
%%% The Bibliography
%%%
%%%%%%%%%%%%%%%%%%%%%%%%%%%%%%%%%%%%%%%%%%%%%%%%%%%%%%%%%%%%%%%%%%%%%%
%%%%%%%%%%%%%%%%%%%%%%%%%%%%%%%%%%%%%%%%%%%%%%%%%%%%%%%%%%%%%%%%%%%%%%

%\bibliographystyle{JHEP.bst}
%\bibliography{negsec}{}

\providecommand{\href}[2]{#2}\begingroup\raggedright\endgroup

%%%%%%%%%%%%%%%%%%%%%%%%%%%%%%%%%%%%%%%%%%%%%%%%%%%%%%%%%%%%%%%%%%%%%%
%%%%%%%%%%%%%%%%%%%%%%%%%%%%%%%%%%%%%%%%%%%%%%%%%%%%%%%%%%%%%%%%%%%%%%
%%%
%%% The End of the Document
%%%
%%%%%%%%%%%%%%%%%%%%%%%%%%%%%%%%%%%%%%%%%%%%%%%%%%%%%%%%%%%%%%%%%%%%%%
%%%%%%%%%%%%%%%%%%%%%%%%%%%%%%%%%%%%%%%%%%%%%%%%%%%%%%%%%%%%%%%%%%%%%%

\end{document}